\documentclass[twocolumn,showpacs,aps,prl,superscriptaddress]{revtex4}
\usepackage{graphicx}
\usepackage{dcolumn}
\usepackage{amsmath}
\usepackage{multirow}
\usepackage{epsfig}
\RequirePackage{xspace}
\def\epem   {\ensuremath{e^+e^-}\xspace}

\def\qqbar  {\ensuremath{q\overline q}\xspace}
\def\piz    {\ensuremath{\pi^0}\xspace}
\def\pip    {\ensuremath{\pi^+}\xspace}
\def\pim    {\ensuremath{\pi^-}\xspace}
\def\pipi   {\ensuremath{\pi^+\pi^-}\xspace}
\def\pipm   {\ensuremath{\pi^\pm}\xspace}

\def\Kbar   {\kern 0.2em\overline{\kern -0.2em K}{}\xspace}

\def\Kz     {\ensuremath{K^0}\xspace}
\def\Kzb    {\ensuremath{\Kbar^0}\xspace}
\def\KzKzb  {\ensuremath{\Kz \kern -0.16em \Kzb}\xspace}
\def\Kp     {\ensuremath{K^+}\xspace}
\def\Km     {\ensuremath{K^-}\xspace}

\def\KpKm   {\ensuremath{\Kp \kern -0.16em \Km}\xspace}
\def\KS     {\ensuremath{K^0_{\scriptscriptstyle S}}\xspace} 
 
\def\Kstarz {\ensuremath{K^{*0}}\xspace}
\def\Kstarzb{\ensuremath{\Kbar^{*0}}\xspace}

\def\Dbar   {\kern 0.2em\overline{\kern -0.2em D}{}\xspace}

\def\Dz     {\ensuremath{D^0}\xspace}
\def\Dzb    {\ensuremath{\Dbar^0}\xspace}
\def\DzDzb  {\ensuremath{\Dz {\kern -0.16em \Dzb}}\xspace}
\def\Dp     {\ensuremath{D^+}\xspace}
\def\Dm     {\ensuremath{D^-}\xspace}

\def\Dmp    {\ensuremath{D^\mp}\xspace}
\def\DpDm   {\ensuremath{\Dp {\kern -0.16em \Dm}}\xspace}

\def\Dstarp {\ensuremath{D^{*+}}\xspace}

\def\Ds     {\ensuremath{D^+_s}\xspace}
\def\Dss    {\ensuremath{D^{*+}_s}\xspace}
\def\Bbar   {\kern 0.18em\overline{\kern -0.18em B}{}\xspace}

\def\BB     {\ensuremath{B\Bbar}\xspace} 
\def\Bz     {\ensuremath{B^0}\xspace}
\def\Bzb    {\ensuremath{\Bbar^0}\xspace}
\def\BzBzb  {\ensuremath{\Bz {\kern -0.16em \Bzb}}\xspace}
\def\Bu     {\ensuremath{B^+}\xspace}
\def\Bub    {\ensuremath{B^-}\xspace}

\def\BpBm   {\ensuremath{\Bu {\kern -0.16em \Bub}}\xspace}
\mathchardef\Upsilon="7107
\def\Y#1S{\ensuremath{\Upsilon{(#1S)}}\xspace}
\def\FourS  {\Y4S}
\def\mds    {\mbox{$M_{D^+_s}$}\xspace}
\def\mbc    {\mbox{$M_{\rm bc}$}\xspace}
\def\DeltaE {\mbox{$\Delta E$}\xspace}
\newcommand{\gev}{{\rm{\,Ge\kern -0.1em V}}\xspace}
\newcommand{\mev}{{\rm{\,Me\kern -0.1em V}}\xspace}
\newcommand{\gevc}{{{{\rm \,Ge\kern -0.1em V\!/}}c}\xspace}
\newcommand{\mevc}{{{{\rm \,Me\kern -0.1em V\!/}}c}\xspace}
\newcommand{\gevcc}{{{{\rm \,Ge\kern -0.1em V\!/}}c^2}\xspace}
\newcommand{\mevcc}{{{{\rm\,Me\kern -0.1em V\!/}}c^2}\xspace}
\def\cm   {\ensuremath{{\rm \,cm}}\xspace}
\def\L{{\ensuremath{\cal L}}\xspace}
\def\to{\ensuremath{\rightarrow}\xspace}
\newcommand{\stat}{\ensuremath{\rm{(stat)}}\xspace}
\newcommand{\syst}{\ensuremath{\rm{(syst)}}\xspace}
\newcommand{\theo}{\ensuremath{\rm{(theo)}}\xspace}

\def\CP {\ensuremath{C\!P}\xspace}
\newcommand{\thetc}{\ensuremath{\theta_{\scriptscriptstyle C}}\xspace}
\newcommand{\figref}[1]{Figure~\ref{fig:#1}}
\newcommand{\tabref}[1]{Table~\ref{tab:#1}}
\renewcommand{\eqref}[1]{Eq.~(\ref{eq:#1})}
\newcommand{\nbb}       {\mbox{$N_{\BB}$}}
\newcommand{\coshel}    {\mbox{$\cos{\theta_{H}}$}}

\def\evtgen{\mbox{\tt EvtGen}\xspace}
\def\geant {\mbox{\tt GEANT3}\xspace}

\newcommand{\bbpairs}   {\mbox{$657\times 10^6$}}
\def\dspiBFal{\ensuremath{\left(1.99\pm0.26\pm0.18\right)\times10^{-5}}}
\def\dskBFal {\ensuremath{\left(1.91\pm0.24\pm0.17\right)\times10^{-5}}}

\begin{document}

\begin{flushleft}
Belle Preprint\,\# 2010-14\\
KEK Preprint\,\# 2010-23
\end{flushleft}

\title{
 {\large 
   \bf Measurements of Branching Fractions for {\boldmath $\Bz\to\Ds\pim$ and $\Bzb\to\Ds\Km$}       
 }
}

\affiliation{Budker Institute of Nuclear Physics, Novosibirsk}
\affiliation{Faculty of Mathematics and Physics, Charles University, Prague}
\affiliation{Chiba University, Chiba}
\affiliation{University of Cincinnati, Cincinnati, Ohio 45221}
\affiliation{Justus-Liebig-Universit\"at Gie\ss{}en, Gie\ss{}en}
\affiliation{The Graduate University for Advanced Studies, Hayama}
\affiliation{Hanyang University, Seoul}
\affiliation{University of Hawaii, Honolulu, Hawaii 96822}
\affiliation{High Energy Accelerator Research Organization (KEK), Tsukuba}
\affiliation{Indian Institute of Technology Guwahati, Guwahati}
\affiliation{Institute of High Energy Physics, Chinese Academy of Sciences, Beijing}
\affiliation{Institute of High Energy Physics, Vienna}
\affiliation{Institute of High Energy Physics, Protvino}
\affiliation{Institute of Mathematical Sciences, Chennai}
\affiliation{Institute for Theoretical and Experimental Physics, Moscow}
\affiliation{J. Stefan Institute, Ljubljana}
\affiliation{Kanagawa University, Yokohama}
\affiliation{Institut f\"ur Experimentelle Kernphysik, Karlsruher Institut f\"ur Technologie, Karlsruhe}
\affiliation{Korea Institute of Science and Technology Information, Daejeon}
\affiliation{Korea University, Seoul}
\affiliation{Kyungpook National University, Taegu}
\affiliation{\'Ecole Polytechnique F\'ed\'erale de Lausanne (EPFL), Lausanne}
\affiliation{Faculty of Mathematics and Physics, University of Ljubljana, Ljubljana}
\affiliation{University of Maribor, Maribor}
\affiliation{Max-Planck-Institut f\"ur Physik, M\"unchen}
\affiliation{University of Melbourne, School of Physics, Victoria 3010}
\affiliation{Nagoya University, Nagoya}
\affiliation{Nara Women's University, Nara}
\affiliation{National Central University, Chung-li}
\affiliation{National United University, Miao Li}
\affiliation{Department of Physics, National Taiwan University, Taipei}
\affiliation{H. Niewodniczanski Institute of Nuclear Physics, Krakow}
\affiliation{Nippon Dental University, Niigata}
\affiliation{Niigata University, Niigata}
\affiliation{Novosibirsk State University, Novosibirsk}
\affiliation{Osaka City University, Osaka}
\affiliation{Panjab University, Chandigarh}
\affiliation{Saga University, Saga}
\affiliation{University of Science and Technology of China, Hefei}
\affiliation{Seoul National University, Seoul}
\affiliation{Sungkyunkwan University, Suwon}
\affiliation{School of Physics, University of Sydney, NSW 2006}
\affiliation{Tata Institute of Fundamental Research, Mumbai}
\affiliation{Excellence Cluster Universe, Technische Universit\"at M\"unchen, Garching}
\affiliation{Tohoku Gakuin University, Tagajo}
\affiliation{Tohoku University, Sendai}
\affiliation{Department of Physics, University of Tokyo, Tokyo}
\affiliation{Tokyo Metropolitan University, Tokyo}
\affiliation{Tokyo University of Agriculture and Technology, Tokyo}
\affiliation{IPNAS, Virginia Polytechnic Institute and State University, Blacksburg, Virginia 24061}
\affiliation{Wayne State University, Detroit, Michigan 48202}
\affiliation{Yonsei University, Seoul}
  \author{A.~Das}\affiliation{Tata Institute of Fundamental Research, Mumbai} 
  \author{T.~Aziz}\affiliation{Tata Institute of Fundamental Research, Mumbai} 
  \author{K.~Trabelsi}\affiliation{High Energy Accelerator Research Organization (KEK), Tsukuba} 
  \author{G.~B.~Mohanty}\affiliation{Tata Institute of Fundamental Research, Mumbai} 
  \author{I.~Adachi}\affiliation{High Energy Accelerator Research Organization (KEK), Tsukuba} 
  \author{H.~Aihara}\affiliation{Department of Physics, University of Tokyo, Tokyo} 
  \author{K.~Arinstein}\affiliation{Budker Institute of Nuclear Physics, Novosibirsk}\affiliation{Novosibirsk State University, Novosibirsk} 
  \author{V.~Aulchenko}\affiliation{Budker Institute of Nuclear Physics, Novosibirsk}\affiliation{Novosibirsk State University, Novosibirsk} 
  \author{T.~Aushev}\affiliation{\'Ecole Polytechnique F\'ed\'erale de Lausanne (EPFL), Lausanne}\affiliation{Institute for Theoretical and Experimental Physics, Moscow} 
  \author{A.~M.~Bakich}\affiliation{School of Physics, University of Sydney, NSW 2006} 
% ...
\collaboration{The Belle Collaboration}

\begin{abstract}
We present improved measurements of the branching fractions for
the decays $\Bz\to\Ds\pim$ and $\Bzb\to\Ds\Km$ using a data sample
of \bbpairs\ \BB\ events collected at the \FourS\ resonance with
the Belle detector at the KEKB asymmetric-energy \epem\ collider.
The results are ${\cal B}\left(\Bz\to\Ds\pim\right)=\dspiBFal$ and
${\cal B}\left(\Bzb\to\Ds\Km\right)=\dskBFal$, where the uncertainties
are statistical and systematic, respectively. Based on these results,
we determine the ratio between amplitudes of the doubly Cabibbo
suppressed decay $\Bz\to\Dp\pim$ and the Cabibbo favored decay
$\Bz\to\Dm\pip$, $R_{D\pi}=\left[1.71\pm0.11\stat\pm0.09\syst
\pm0.02\theo\right]\%$, where the last term denotes the theory error.
\end{abstract}


\maketitle

In the standard model (SM), \CP\ violation occurs due to a single
irreducible phase appearing in the quark-flavor mixing matrix,
called the Cabibbo-Kobayashi-Maskawa (CKM) matrix~\cite{ckm},
which relates the weak interaction eigenstates to those of mass.
Unitarity of the CKM matrix yields relationships between its
elements that can be depicted as triangles in the complex plane~\cite{ut}.
$B$ meson decays offer a variety of ways to measure the angles
and sides of the unitarity triangle (UT), formed from elements
in the first and third columns of the CKM matrix, and, hence to
verify the \CP\ violation mechanism of the SM.

\begin{figure}[hbtp]
 \vspace{4pt}
  \hbox{
   \hspace{0.15in}
    \epsfxsize=1.4in
    \epsffile{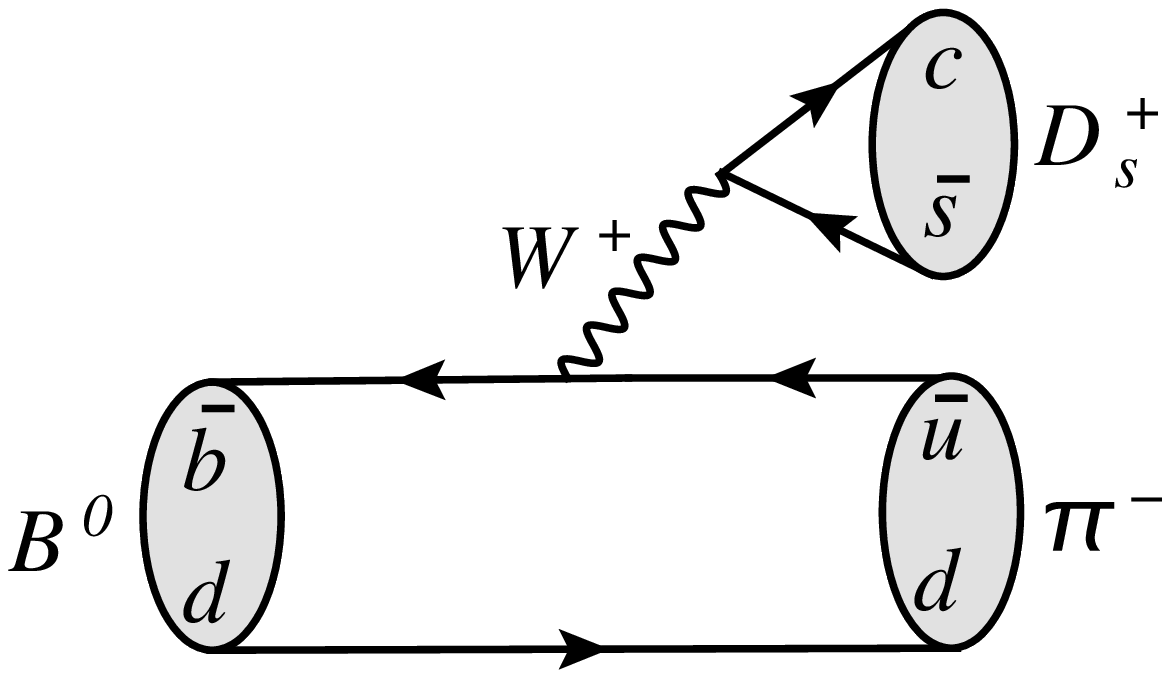}
    \hspace{0.0007in}
    \epsfxsize=1.4in
    \epsffile{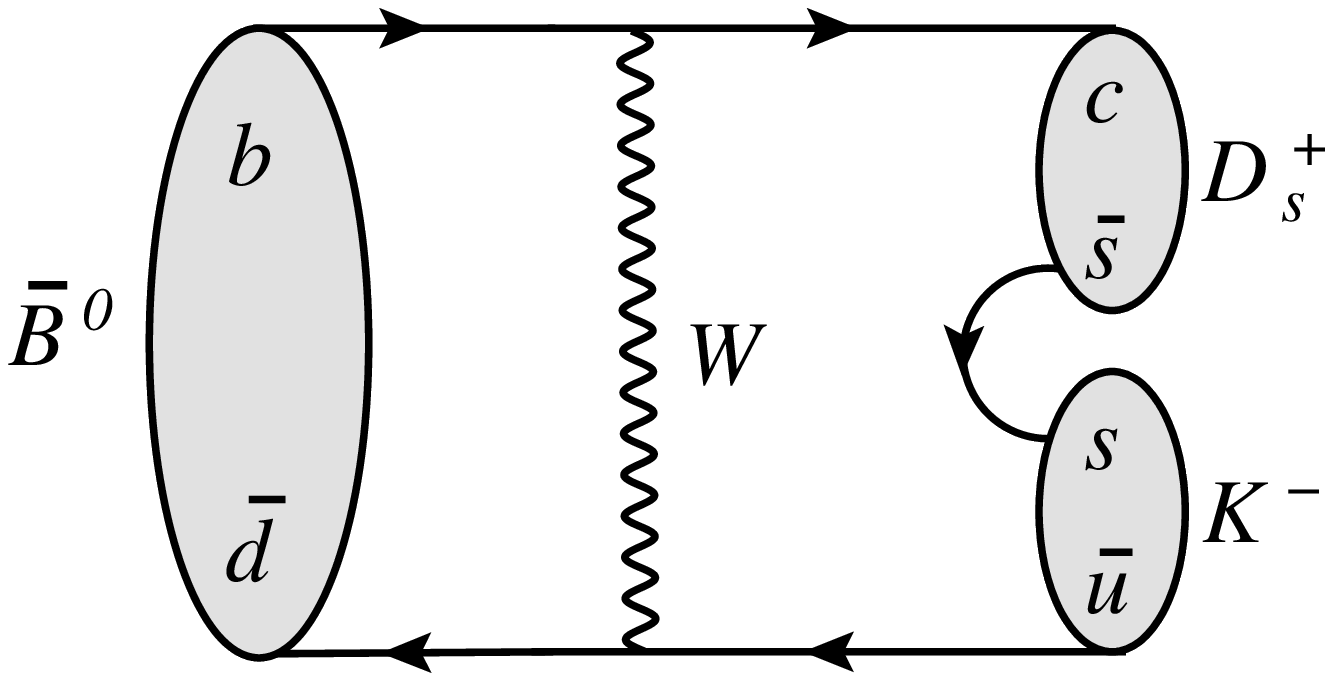}
    }
 \vspace{8pt}
  \hbox{ 
   \hspace{0.15in}
    \epsfxsize=1.4in
    \epsffile{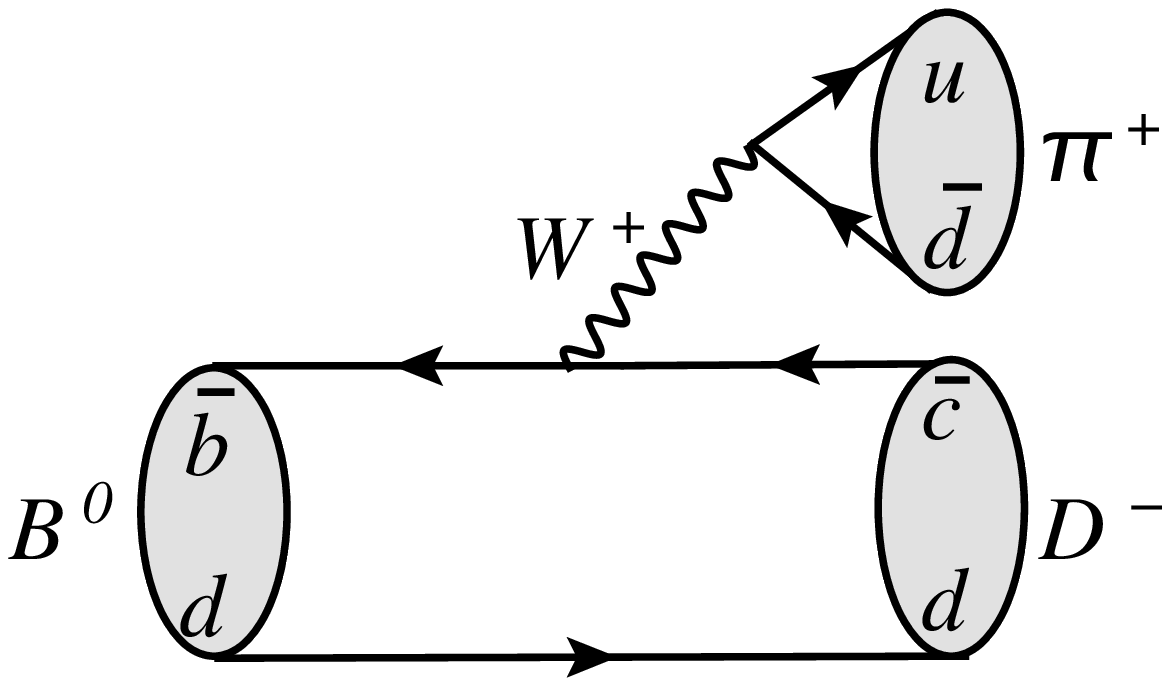}
    \hspace{0.0007in}
    \epsfxsize=1.4in
    \epsffile{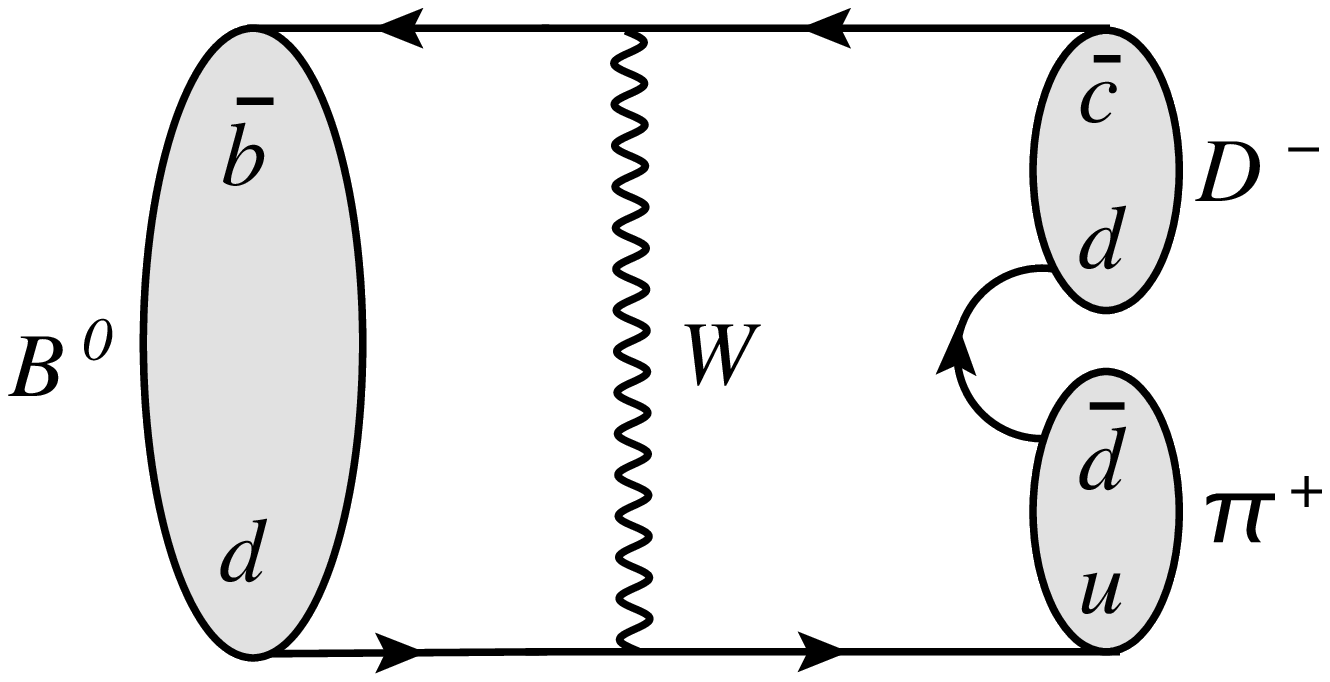}
    }
 \vspace{8pt}
  \hbox{ 
   \hspace{0.15in}
    \epsfxsize=1.4in
    \epsffile{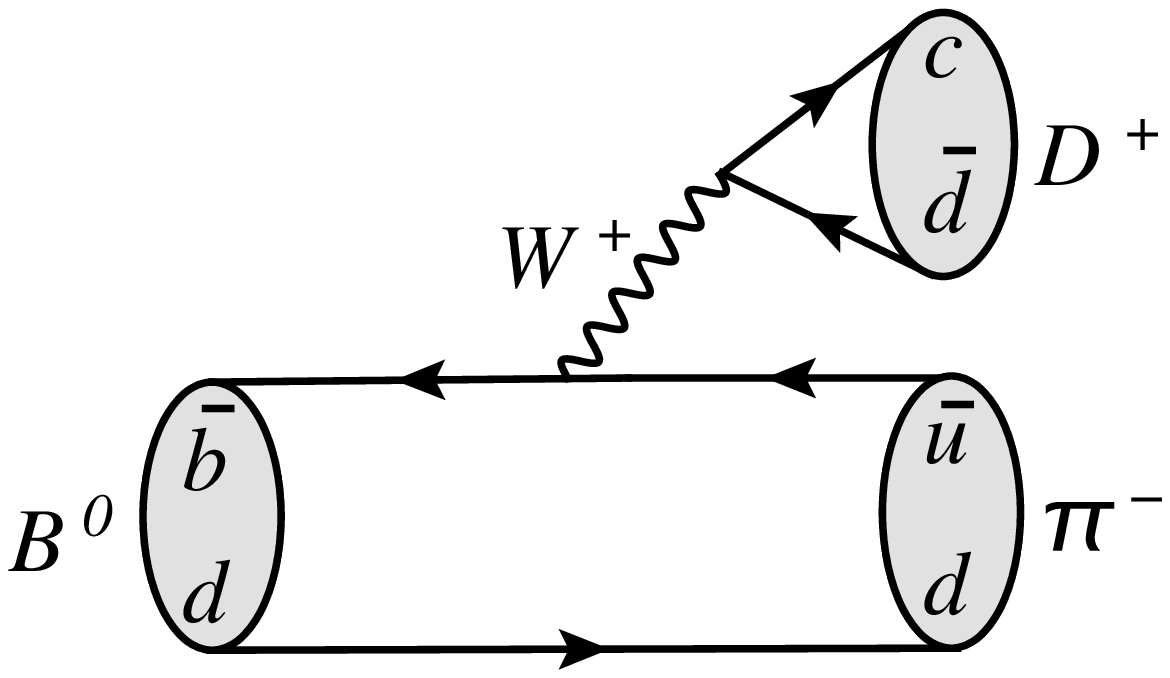}
    \hspace{0.0007in}
    \epsfxsize=1.4in 
    \epsffile{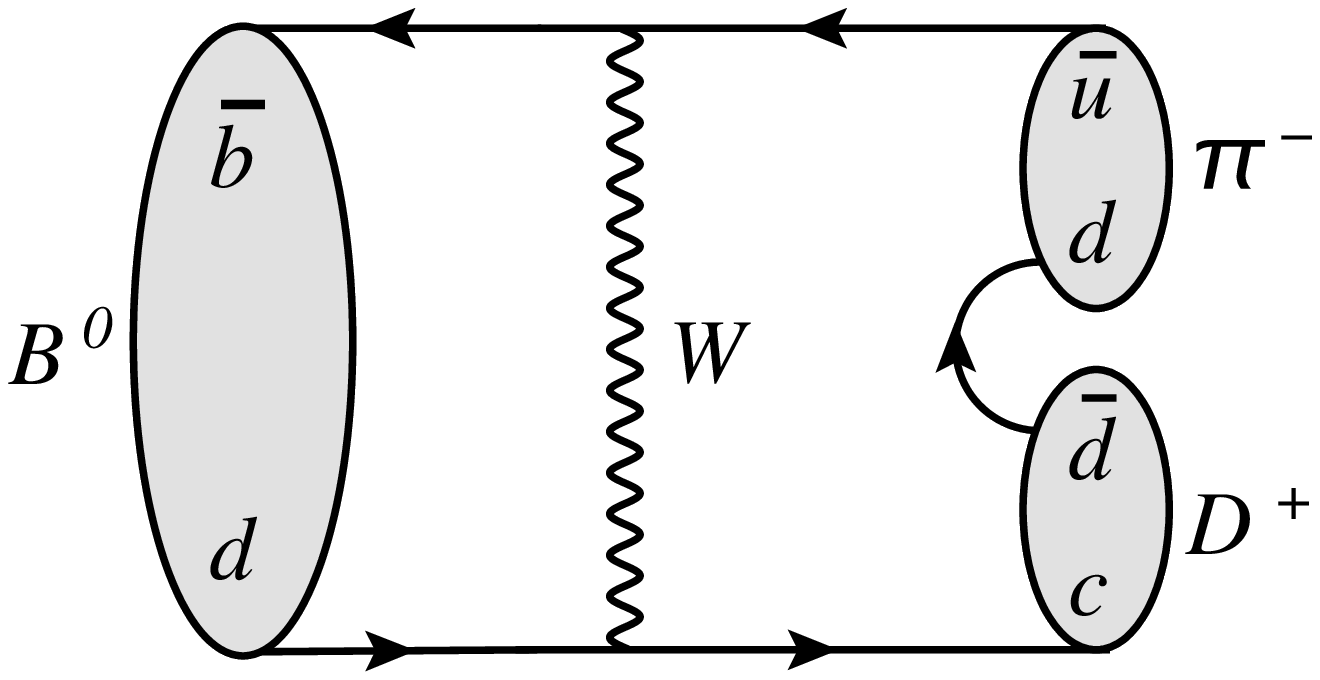}
    }
 \hbox{                 \hspace{0.8in} (c)\hspace{1.3 in} (f)}
  \vskip -2.75 cm \hbox{\hspace{0.8in} (b)\hspace{1.3 in} (e)}
  \vskip -2.75 cm \hbox{\hspace{0.8in} (a)\hspace{1.3 in} (d)}
  \vskip 4.65cm
 \caption{Feynman diagrams for (a) $\Bz\to\Ds\pim$; (b) the Cabibbo
 favored decay $\Bz\to\Dm\pip$; (c) the doubly Cabibbo suppressed decay
 $\Bz\to\Dp\pim$; and the color suppressed $W$-exchange contribution
 to (d) $\Bzb\to\Ds\Km$, (e) $\Bz\to\Dm\pip$, and (f) $\Bz\to\Dp\pim$.}
 \label{fig:fey1}
\end{figure}

Of particular interest is the decay $\Bz\to\Ds\pim$, which is
dominated by the tree level $b\to u$ transition shown in
Fig.~\ref{fig:fey1}(a). Assuming SU(3) flavor symmetry, one can
use this decay channel to determine the ratio between amplitudes
of the doubly Cabibbo suppressed decay $\Bz\to\Dp\pim$ and the
Cabibbo favored decay $\Bz\to\Dm\pip$~\cite{rdpi}
\begin{equation}
 \label{eq:rdpi}
 R_{D\pi}=\tan\thetc\, \frac{f_D}{f_{D_s}}\,
 \sqrt{\frac{{\cal B}(\Bz\to\Ds\pim)}{{\cal B}(\Bz\to\Dm\pip)}},
\end{equation}
where $\thetc$ is the Cabibbo angle, $f_{D}$\,$(f_{D_s})$ is the
$D$\,$(D_s)$ meson decay constant, and ${\cal B}(\Bz\to\Ds\pim)$ and
${\cal B}(\Bz\to\Dm\pip)$ are the branching fractions of $\Bz\to\Ds\pim$
and $\Bz\to\Dm\pip$. In Figs.~\ref{fig:fey1}(b) and \ref{fig:fey1}(c) we
show the dominant Feynman diagrams for the decays $\Bz\to\Dmp\pipm$. The
ratio $R_{D\pi}$ is an important input for the determination of the UT
angle $\phi_3$, since the measurement of time-dependent $\CP$ violation
in $\Bz\to\Dmp\pipm$~\cite{dpicpv} determines only the quantity
$R_{D\pi}\sin(2\phi_1+\phi_3)/(1+R_{D\pi}^{\,2})$, where $\phi_1$ is
the most precisely measured angle of the UT~\cite{bellephi1,babarbeta}.
Furthermore, it has also been suggested~\cite{vub} that the CKM matrix
element $|V_{ub}|$ (related to one of the sides of the UT) can be
extracted from the measured branching fraction of $\Bz\to\Ds\pim$.

The decay $\Bzb\to\Ds\Km$ occurs via the internal $W$-exchange
diagram [see Fig.~\ref{fig:fey1}(d)]. Potential contributions
arising from rescattering effects~\cite{dsk1} could enhance its
branching fraction. Recent studies~\cite{dsk2}, however, find
the rescattering contribution to be negligible. Furthermore,
the calculation of $R_{D\pi}$ in Ref.~\cite{rdpi} assumes that
the size of the $W$-exchange amplitude in $\Bz\to\Dmp\pipm$
[Figs.~\ref{fig:fey1}(e) and \ref{fig:fey1}(f)] is small compared
to the corresponding tree amplitude. One can verify this hypothesis
with an accurate measurement of ${\cal B}(\Bzb\to\Ds\Km)$. In
the absence of rescattering the exchange diagram is the sole
contributor to $\Bzb\to\Ds\Km$, and hence it provides a measure
of the $W$-exchange contribution in $\Bz\to\Dmp\pipm$.

The decay channels $\Bz\to\Ds\pim$ and $\Bzb\to\Ds\Km$ have been
previously studied by the Belle~\cite{belleold} and BaBar~\cite{babarold}
collaborations. In this paper, we present improved measurements of the
branching fractions ${\cal B}(\Bz\to\Ds\pim)$ and ${\cal B}(\Bzb\to\Ds\Km)$
based on a data sample of $\bbpairs$ $\BB$ decays, which is close to $8$
times the size of the one used in our earlier result~\cite{belleold}. The
data were collected with the Belle detector at the KEKB asymmetric-energy
$\epem$ collider~\cite{kekb}. A detailed description of the Belle detector
can be found elsewhere~\cite{belledetector}.

We select $\Bz\to\Ds\pim$ and $\Bzb\to\Ds\Km$ decay candidates~\cite{charge}
from events that have four or more charged tracks. Each track is required
to be well measured in a tracking system that consists of a silicon
vertex detector and a central drift chamber (both operating in a 1.5\,T
magnetic field), and to originate from the interaction point (IP).
Track candidates must have a minimum transverse momentum of $100\mevc$,
and a distance of closest approach with respect to the IP less than
$0.2\cm$ in the $r$--$\phi$ plane, which is perpendicular to the $z$
axis, and less than $4.0\cm$ along the $z$ axis, where the $z$ axis
is defined by the direction opposite to the $e^+$ beam. Charged pions
and kaons are identified by combining particle identification (PID)
information obtained with various subdetectors: ionization energy loss
from the drift chamber, time-of-flight information from an array of
scintillation counters, and the number of photoelectrons from an aerogel
Cherenkov counter system. We distinguish kaons from pions using a
likelihood ratio, ${\cal R}_{K/\pi}=\L_K/(\L_K+\L_\pi)$, where $\L_K$
$(\L_\pi)$ is the likelihood value for the kaon (pion) hypothesis. We
require ${\cal R}_{K/\pi}$ to be greater than $0.6$ for kaon candidates,
while tracks failing this requirement are classified as pions. The
efficiency for kaon (pion) identification ranges between $84\%$ to
$98\%$ ($92\%$ to $94\%$) depending on the track momentum with a
pion (kaon) fake rate of about $8\%$ ($16\%$).

We reconstruct $\Ds$ mesons in three decay modes: $\Ds\to\phi\pip$,
$\Kstarzb\Kp$, and $\KS\Kp$. The $\phi$ $(\Kstarzb)$ mesons are
formed from $\Kp\Km$ $(\Km\pip)$ pairs having invariant masses
that lie within $14\mevcc$ $(75\mevcc)$ of the nominal $\phi$
$(\Kstarzb)$ mass~\cite{PDG}. Note that for kaons originating
from a $\phi$ decay we relax the ${\cal R}_{K/\pi}$ requirement
to $0.1$ due to the small background contribution. To reduce
combinatorial background, we require $|\coshel|>0.3$ for the
$\Ds\to\phi\pip$ ($\Ds\to\Kstarzb\Kp$) mode, where $\theta_H$ is
the angle between decay products of the $\phi$ ($\Kstarzb$) and
the flight direction of the $\Ds$ meson in the rest frame of
the $\phi$ ($\Kstarzb$). We reconstruct $\KS$ mesons through the
channel $\KS\to\pipi$, where we require the invariant mass of
two oppositely charged tracks (with the pion mass hypothesis
assumed) to be within $10\mevcc$ of the nominal $\KS$ mass~\cite{PDG}.
The $\KS$ candidates must also satisfy momentum-dependent selection
criteria based on their vertex topology and flight length in the
$r$--$\phi$ plane~\cite{selks}. We select $\Ds$ mesons in a wide
mass window ($1.92\gevcc <\mds <2.02\gevcc$), common to the three
decay modes, for further studies. Finally we combine each $\Ds$
candidate with an oppositely charged pion or kaon to form a
neutral $B$ meson.

For the reconstruction of $B$ candidates we utilize two kinematic
variables: the center-of-mass (CM) energy difference, $\DeltaE=E_B-
E_{\rm beam}$, and the beam-constrained mass, $\mbc=\sqrt{E_{\rm beam}^2
-\vec{p}_B^{\,2}}$, where $E_{\rm beam}$ is the beam energy, $E_B$
and $\vec{p}_B$ are the energy and momentum of the $B$ candidate
measured in the CM frame, respectively ($c=1$ is assumed). The $\mbc$
distribution for signal events peaks near the $B$ mass, while the
$\DeltaE$ distribution peaks at zero. We retain $B$ candidates with
$5.27\gevcc <\mbc <5.30\gevcc$ and $-0.1\gev <\DeltaE <0.2\gev$. An
asymmetric $\DeltaE$ requirement is imposed to suppress background
contributions from $B$ decays, such as $\Bz\to\Dss\pim$ and
$\Bzb\to\Dss\Km$, at negative $\DeltaE$ values.

About $5\%$ of the selected $\Bz\to\Ds\pim$ and $\Bzb\to\Ds\Km$ events
contain multiple $B$ candidates. In such cases we choose the one with
the $\mbc$ value closest to the nominal $\Bz$ mass~\cite{PDG}. In
order to determine the background reduction criteria (described below),
signal and background yields are estimated in the signal region with
the requirements that $|\DeltaE|<30\mev$ and that $\mds$ be within
$13\mevcc$, $15\mevcc$, and $17\mevcc$ of the nominal $\Ds$ mass for
$\phi\pip$, $\Kstarzb\Kp$, and $\KS\Kp$, respectively. These requirements
roughly correspond to a $\pm 3\sigma$ window in resolution.

Continuum $\epem\to\qqbar$ ($q=u,\,d,\,s$, and $c$ quarks) events
are the dominant background. To discriminate the jet-like continuum
background from signal we use modified Fox-Wolfram moments~\cite{fox}
that are combined into a Fisher discriminant. We further combine
the Fisher output with the cosine of the angle between the $B$ flight
direction in the CM frame and the $z$ axis, to form a likelihood ratio
${\cal R}=\L_{\rm sig}/(\L_{\rm sig}+\L_{\qqbar})$. Here, $\L_{\rm sig}$
and $\L_{\qqbar}$ are the likelihood functions for signal and continuum
events obtained with Monte Carlo (MC) simulations~\cite{MCsim}. We
impose separate requirements on ${\cal R}$ for the three decay modes
in both $\Bz\to\Ds\pim$ and $\Bzb\to\Ds\Km$. These requirements
are obtained by maximizing a figure of merit, $S/\sqrt{S+B}$, where
$S$ and $B$ are the number of signal and $\qqbar$ events expected
in the signal region, calculated using MC simulated events. The
requirements on ${\cal R}$ remove $92\%$ ($78\%$) of continuum
background while retaining $75\%$ ($86\%$) of signal events for
$\Bz\to\Ds\pim$ ($\Bzb\to\Ds\Km$).

A large MC sample of $\BB$ events is used to determine possible
backgrounds that can contaminate our signal region. The decay
$\Bzb\to\Dp\pim$, $\Dp\to\Km\pip\pip$ including $\Dp\to\Kstarzb\pip$
and $\Kstarz_0(1430)\pip$, where a pion is misidentified as a kaon,
poses a particular challenge for the $\Bz\to\Ds\pim$ channel. This
decay mode has a large branching fraction; its reconstructed invariant
mass spectrum peaks near the $\Ds$ peak while its $\DeltaE$ distribution
is shifted by about $70\mev$ from zero. The $\Bzb\to\Dp\pim$ background
is more prominent in $\Ds\to\Kstarzb\Kp$ compared to $\Ds\to\phi\pip$
because of the wider invariant-mass requirement. To suppress this
background, we reject event candidates that are consistent with the
$\Dp\to\Km\pip\pip$ mass hypothesis within $16\mevcc$ ($\sim 3\sigma$)
when the two same-sign particles in the $\Ds$ candidate are assigned
to be pions. For the $\Ds\to\KS\Kp$ mode there is a similar background
from $\Bzb\to\Dp\pim$, $\Dp\to\KS\pip$. Here we exclude candidates
consistent within $20\mevcc$ with the $\Dp\to\KS\pip$ mass hypothesis.
The channel $\Bzb\to\Ds\Km$ has a similar reflection background from
$\Bzb\to\Dp\Km$, $\Dp\to\Km\pip\pip$. We apply the same rejection
criteria, as in $\Bz\to\Ds\pim$, to the invariant mass of the
$\Km\pip\pip$ system. Our invariant-mass veto requirements also reduce
a similar background from $\Bzb\to\Dstarp\pim$, $\Dstarp\to\Dp\piz$. 

Another $\BB$ background arises from charmless decays such as
$\Bz\to\KS\Km\pip$, $\KS\Kp\Km$, $\Kstarzb\Kp\Km$, and $\phi\Km\pip$.
These events peak at $\DeltaE=0$ as the final state is the same as
signal, but have a broad nonpeaking $\Ds$ mass distribution due to
the absence of a $\Ds$ in the final state. Finally, there is a
crossfeed contribution from $\Bzb\to\Ds\Km$ ($\Bz\to\Ds\pim$) to
$\Bz\to\Ds\pim$ ($\Bzb\to\Ds\Km$) due to a kaon (pion) faking a pion
(kaon), which also needs to be considered.

To determine the branching fractions of $\Bz\to\Ds\pim$ and
$\Bzb\to\Ds\Km$, we perform an unbinned extended maximum likelihood
fit to the candidate events found in the selected regions of $\mbc$,
$\DeltaE$, and $\mds$ (described above). The probability density
functions (PDFs) are functions of $\DeltaE$ and $\mds$. The extended
likelihood function is
\begin{equation}
\label{eq:eml}
\L =
\frac{e^{-\left(\displaystyle\sum_{j,m}Y_{jm}\right)}}{N!}\displaystyle\prod_{i=1}^{N}
\Bigl\{\displaystyle\sum_{j}Y_{jm}{\cal P}_{jm}(\vec{x}_i)\Bigr\},
\end{equation}
where $Y_{jm}$ is the yield of event category $j$ for $\Ds$ decay mode
$m$ ($\phi\pip$, $\Kstarzb\Kp$, or $\KS\Kp$), $N$ is the total number
of candidate events in three $\Ds$ modes, and ${\cal P}_{jm}(\vec{x}_i)$
is the PDF evaluated for the variables $\vec{x}\equiv (\DeltaE,\mds)$
measured for event $i$. To constrain the three $\Ds$ modes to have a
common branching fraction, we express the signal ($j=1$) yield as
\begin{equation}
 Y_{1m}=\nbb \,{\cal B}\,{\cal B}_m\,\epsilon_m,
\end{equation}
where $\nbb$ is the number of $\BB$ events, ${\cal B}$ is the branching
fraction of $\Bz\to\Ds\pim$ (or, $\Bzb\to\Ds\Km$), ${\cal B}_m$ is the
branching fraction of the $\Ds$ decay mode $m$~\cite{PDG}, and $\epsilon_m$
is the detection efficiency of the corresponding decay mode. Finally to
account for crossfeeds between the two signal channels, they are simultaneously
fitted, with the $\Bzb\to\Ds\Km$ signal yield in the correctly reconstructed
sample determining the normalization of the crossfeed in the $\Bz\to\Ds\pim$
fit region, and vice versa.

There are four PDF components, each denoting an event
category, for the $\Ds$ decay modes considered: signal,
crossfeed, combinatorial, and charmless backgrounds. The signal
$\DeltaE$ PDF shape is modeled with the sum of two Gaussian
functions with a common ratio of the narrow component to the
total for the three $\Ds$ modes. We parametrize the signal
$\mds$ distribution using the sum of two Gaussians with a
common mean and ratio of areas for all the $\Ds$ modes, and use
the same PDF for both the $\Ds\pim$ and $\Ds\Km$ channels. We
use an asymmetric Gaussian to model the $\DeltaE$ distribution
of the $\Ds\pim$ ($\Ds\Km$) crossfeed, that contributes to the
$\Bzb\to\Ds\Km$ ($\Bz\to\Ds\pim$) signal. Combinatorial
background arises when a random track is combined with a
correctly reconstructed or misreconstructed $\Ds$ candidate.
This background is mostly from generic $\BB$ and continuum
$\qqbar$ processes. To model misreconstructed $\Ds$ candidates
we use a linear function to describe the $\mds$ distribution,
while the signal $\mds$ PDF shape is used for combinatorial
background that contains correctly reconstructed $\Ds$
candidates. The $\DeltaE$ distribution in both cases is
parametrized with a linear function. Charmless background
events are characterized by a linear $\mds$ distribution
and a peaking $\DeltaE$, which is modeled with the signal
PDF shape. For both signal and background PDF parametrizations,
we obtain shape parameters from the corresponding MC samples.

We calibrate various PDF shape parameters obtained from
MC events using a large-statistics control sample of
$\Bzb\to\Dp\pim$; $\Dp\to\phi\pip$, $\Kstarzb\Kp$, and
$\KS\Kp$. The peak positions and widths are adjusted
based on the difference between data and MC simulations
observed in the control channel. We find the measured
branching fraction of the control sample is in agreement
with the current world-average value. We also cross-check
our analysis procedure by applying it to a data sample
enriched with the Cabibbo suppressed decay $\Bzb\to\Dp\Km$,
where the $\Dp$ decays to $\phi\pip$, $\Kstarzb\Kp$, and
$\KS\Kp$. The measured branching fraction is found to be
consistent with the world-average value.

\begin{figure*}[!htb]
\center
 \vspace{6pt}
  \hbox{ 
   \hspace{0.0in}
   \epsfxsize=3.2in
   \epsffile{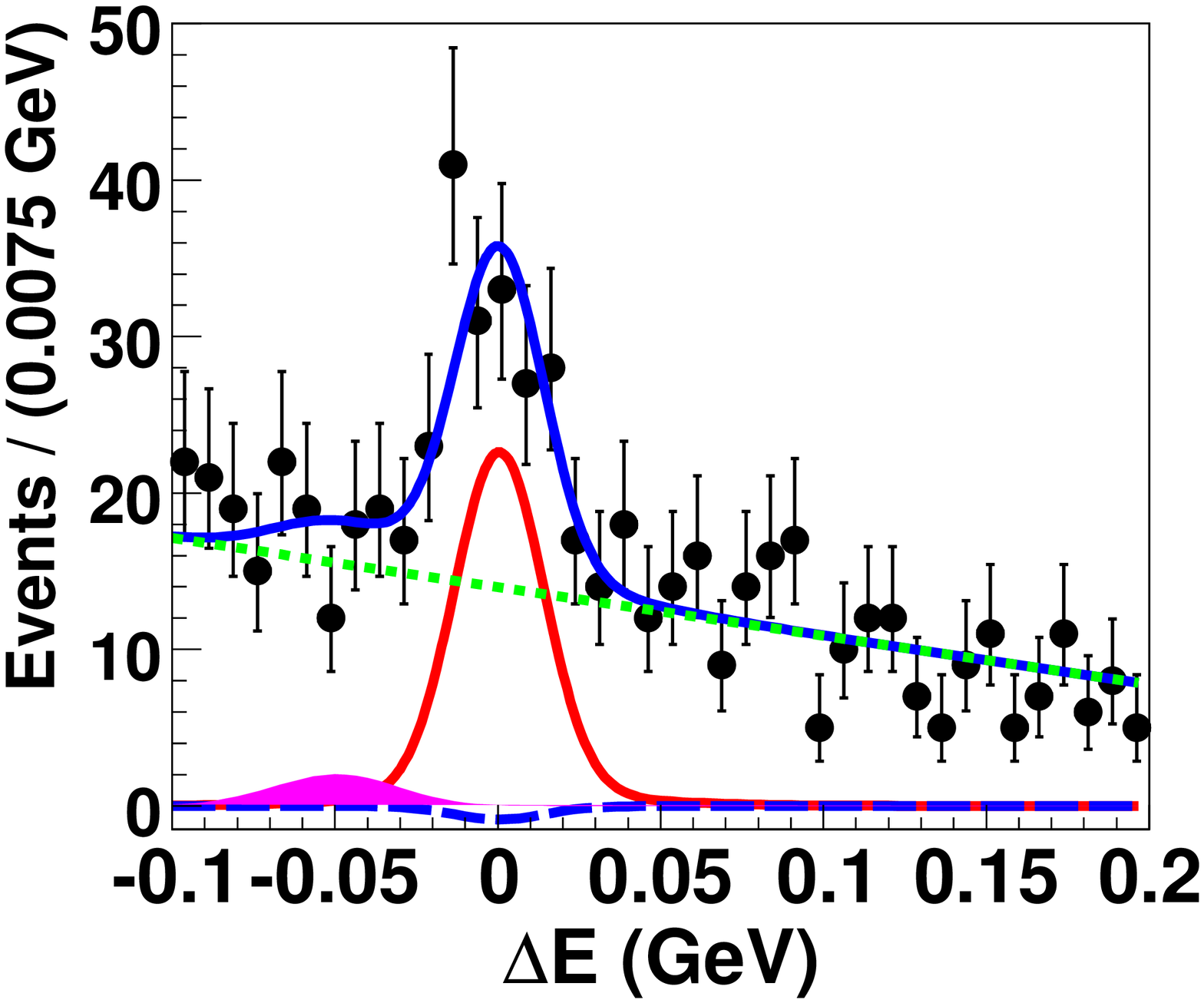}
   \hspace{0.0007in}
   \epsfxsize=3.2in
   \epsffile{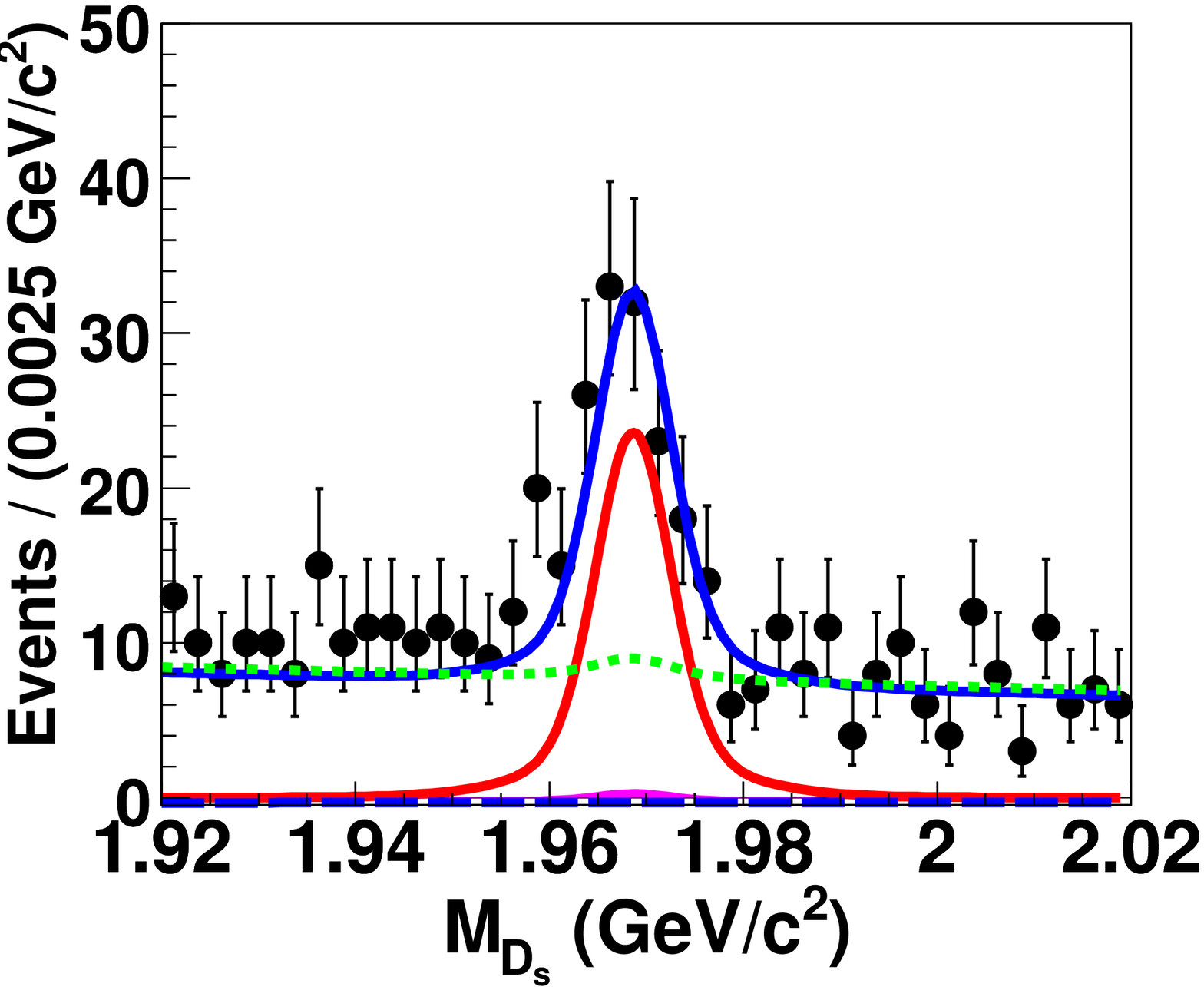}
   }
 \vspace{6pt}
  \hbox{ 
   \hspace{0.0in}
   \epsfxsize=3.2in
   \epsffile{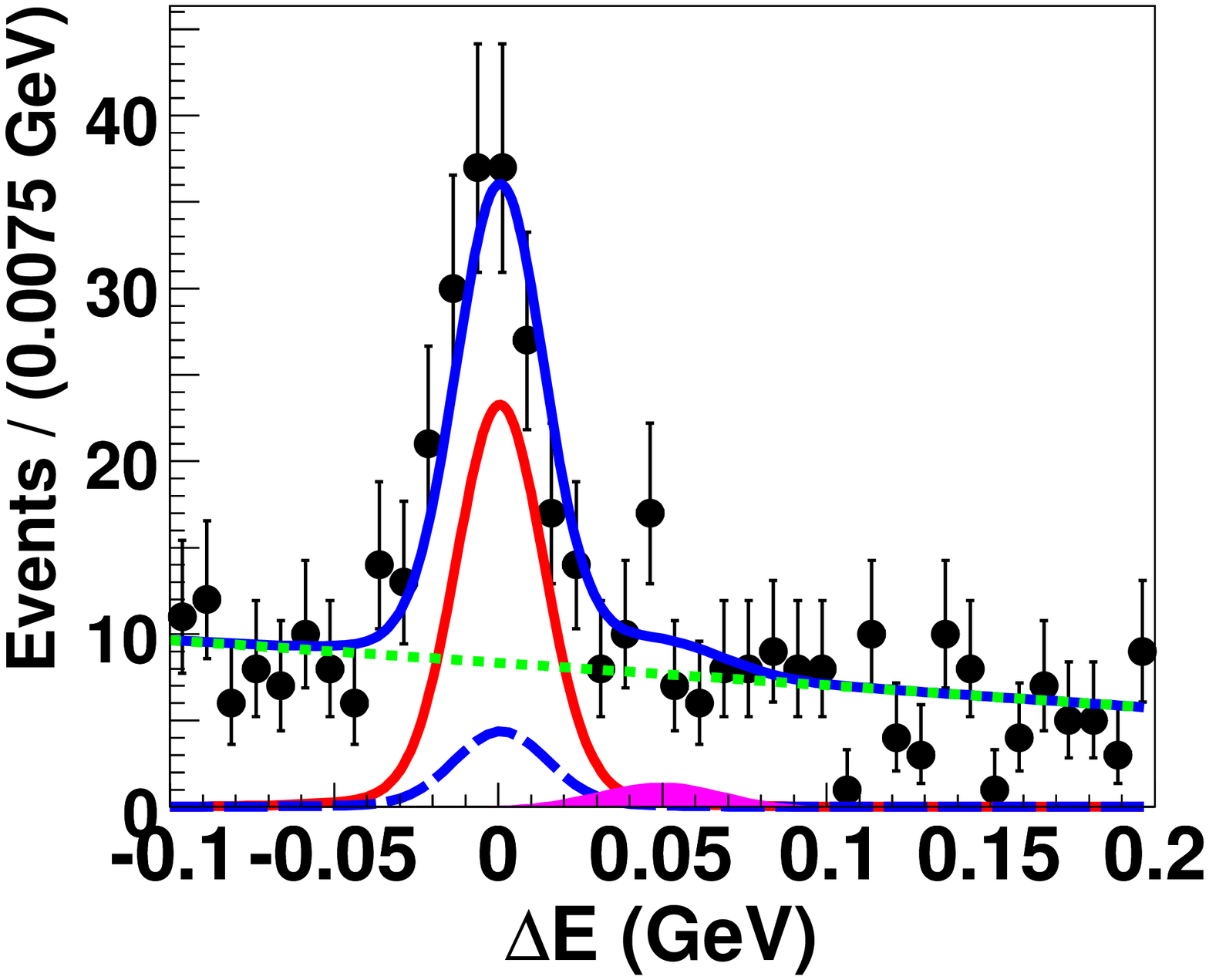}
   \hspace{0.0007in}
   \epsfxsize=3.2in 
   \epsffile{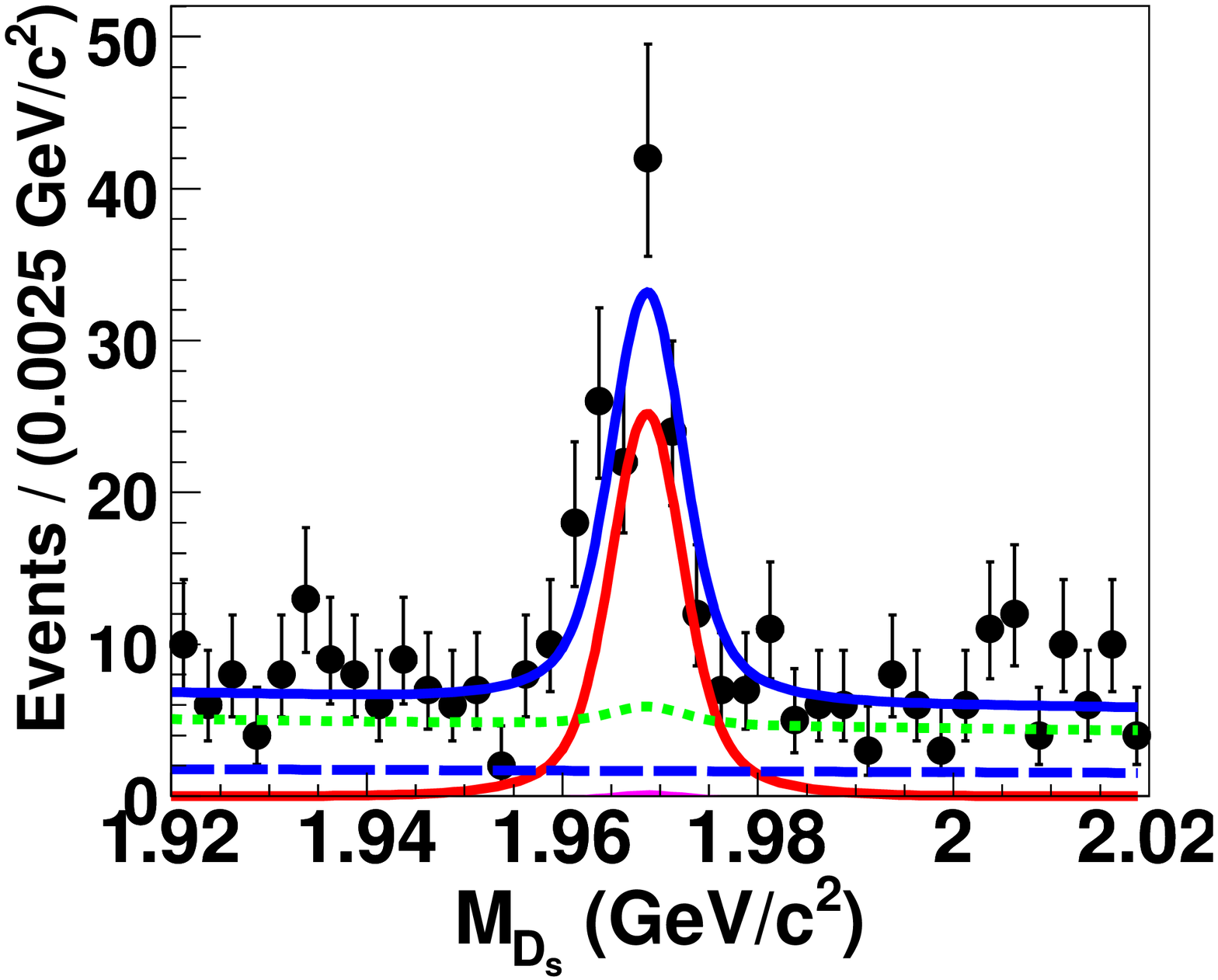}
   }
  \vskip -6.0 cm \hbox{\hspace{0.7in} (b)\hspace{3.0 in} (d)}
  \vskip -7.3 cm \hbox{\hspace{0.7in} (a)\hspace{3.0 in} (c)}
  \vskip 12.4cm
  \caption{(color online).
  Projections of the simultaneous fit to (a,c) $\Bz\to\Ds\pim$
  and (b,d) $\Bzb\to\Ds\Km$. (a,b) correspond to $\DeltaE$ and
  (c,d) are the $\Ds$ mass distributions. Points with error bars
  show the data, the blue solid curves are the total fit result,
  the red solid curves are the signal component, the magenta
  filled curves represent the crossfeed contribution, the green
  dotted curves are the combinatorial background, and the blue
  dashed curves correspond to the charmless $B$ background.
}
\label{fig:sig-proj}
\end{figure*}

Our fit in total has $32$ free parameters. They are the
branching fractions of both the decay channels ($2$); the yields
of the charmless background ($6$), the combinatorial background
with correctly reconstructed $\Ds$ candidates ($6$), and the
pure combinatorial background ($6$); and the slopes of the
linear functions representing the nonpeaking $\DeltaE$ shape
($6$), and the nonpeaking $\mds$ shape ($6$). \figref{sig-proj}
shows results of the simultaneous fit for both the signal
channels, projected onto $\DeltaE$ and $\mds$. For $\DeltaE$
($\mds$) projections we apply the $\mds$ ($\DeltaE$) signal
region requirement, as described earlier. In \tabref{result}
we summarize the fit results. The signal significance is
calculated as $\sqrt{-2\ln({\cal L}_0/{\cal L}_{\rm max})}$,
where ${\cal L}_{\rm max}$ is the maximum likelihood value for
the nominal data fit, and ${\cal L}_0$ is the corresponding value
with the signal yield fixed to zero. Including systematic errors
(described below), which impact only the signal yield, into the
statistical likelihood curve, through a Gaussian convolution,
we determine the significance to be $8.0$ and $9.2$ standard
deviations for $\Bz\to\Ds\pim$ and $\Bzb\to\Ds\Km$, respectively.

\begin{table*}[htb]
\caption{Efficiency ($\epsilon$), signal yield $(N_{\rm sig})$,
 charmless background yield $(N_{\rm chmls})$, and branching fraction
 $({\cal B})$ from fits to the data obtained individually in the three
 $\Ds$ modes as well as from the simultaneous fit. Individual branching
 fraction results (statistical errors only) are consistent with each
 other and with that from the simultaneous fit, where the systematic
 error and signal significance (${\cal S}$) are also quoted.}
\centering
\begin{tabular}{llp{0.0cm}cp{0.3cm}cp{0.3cm}cp{0.1cm}cp{0.3cm}c}
\hline\hline
$B$ mode&$\Ds$ mode&&$\epsilon\,(\%)$&&$N_{\rm sig}$&&$N_{\rm chmls}$&&\multicolumn{1}{c}{${\cal B}\,(10^{-5})$}&&${\cal S}\,(\sigma)$\\
\hline
\multirow{4}{*}{$\Bz\to\Ds\pim$} 
& $\phi(\Kp\Km)\pip$    && $21.6$ && $64\pm 10$ && $0\pm  8$ && $2.08\pm 0.34$ && \\
& $\Kstarzb(\Km\pip)\Kp$&& $11.2$ && $33\pm 9$  && $-7\pm 17$&& $1.71\pm 0.49$ && \\
& $\KS\Kp$              && $15.7$ && $24\pm 9$  && $-4\pm 13$&& $2.21\pm 0.83$ && \\
\cline{2-12}
&\multicolumn{5}{l}{Simultaneous fit result} && && $1.99\pm 0.26\pm 0.18$ && $8.0$\\
\hline
\multirow{4}{*}{$\Bzb\to\Ds\Km$} 
& $\phi(\Kp\Km)\pip$    && $22.0$ && $61\pm 10$ && $14\pm 10$&& $1.97\pm 0.31$ && \\
& $\Kstarzb(\Km\pip)\Kp$&& $11.1$ && $39\pm 9$  && $27\pm 14$&& $2.04\pm 0.47$ && \\
& $\KS\Kp$              && $14.9$ && $19\pm 11$ && $31\pm 12$&& $1.20\pm 0.68$ && \\
\cline{2-12}
&\multicolumn{5}{l}{Simultaneous fit result} && && $1.91\pm 0.24\pm 0.17$ && $9.2$\\
\hline\hline
\end{tabular}
\label{tab:result}
\end{table*}

Systematic uncertainties that affect our measurement are summarized
in \tabref{systematic}. The dominant one is the error on the current
world-average values of the $\Ds$ decay branching fractions~\cite{PDG}.
The remaining sources of systematic error are the fixed PDF shapes, for
which we vary the correction factors (applied to the peak positions
and widths) in accordance with their errors obtained from the control
sample $\Bzb\to\Dp\pim$; MC statistics; the efficiencies of tracking,
PID, and $\KS$ reconstruction; the error on $\nbb$, assuming equal
production of $\BzBzb$ and $\BpBm$ pairs at the $\FourS$; requirements
on ${\cal R}$, evaluated using the control sample; and the fit bias.
We estimate the systematic error due to fit bias as a linear sum of
the bias itself and the statistical error on it, using ensembles of
simulated experiments.

\begin{table}[htb]
\caption{Summary of the systematic uncertainty.}
\centering
\begin{tabular}{lcccc}
\hline\hline
Source & & &\multicolumn{2}{c}{Systematic contribution (\%) to}\\
       & & &${\cal B}(\Bz\to\Ds\pim)$&${\cal B}(\Bzb\to\Ds\Km)$\\
\hline
$\Ds$ branching fraction   & & &$+6.59,-6.51$&$+6.31,-6.14$\\
\hline
PDF shape                  & & &$+1.44,-1.79$&$+1.28,-1.33$\\
MC statistics              & & &$+0.39,-0.48$&$+0.46,-0.45$\\
$\KS$ reconstruction       & & &$+0.45,-0.40$&$\pm 0.63$   \\
PID efficiency             & & &$\pm 2.78$   &$\pm 3.06$   \\
Tracking efficiency        & & &$\pm 4.00$   &$\pm 4.00$   \\
$N_{\BB}$                  & & &$\pm 1.40$   &$\pm 1.40$   \\
Requirements on ${\cal R}$ & & &$\pm 1.60$   &$\pm 1.60$   \\
Fit bias                   & & &$\pm 3.42$   &$\pm 2.09$   \\
\hline
Total                      & & &$+9.26,-9.27$&$+8.74,-8.62$\\
\hline
\end{tabular}
\label{tab:systematic}
\end{table}

We obtain the branching fractions ${\cal B}\left(\Bz\to\Ds\pim\right)=\dspiBFal$
and ${\cal B}\left(\Bzb\to\Ds\Km\right)=\dskBFal$, where the uncertainties
are statistical and systematic, respectively. These results are consistent
with, and constitute a significant improvement over, our previous
results~\cite{belleold}. Using our measurement of $\Bz\to\Ds\pim$
in conjunction with the value of Cabibbo angle~\cite{PDG},
$\tan\thetc=0.2314\pm0.0021$, the lattice QCD calculation of
$f_{D_s}/f_D=1.164\pm0.011$~\cite{dsratio}, and the branching fraction
${\cal B}(\Bz\to\Dm\pip)=(2.68\pm0.13)\times10^{-3}$~\cite{PDG}, we obtain
$R_{D\pi}=[1.71\pm0.11\stat\pm0.09\syst\pm0.02\theo]\%$, where the last term
accounts for the theory uncertainty in the $f_{D}/f_{D_s}$ estimation.
Uncertainties due to other possible SU(3) breaking effects~\cite{maxbaak},
which are of order $(10$-$15)\%$, are not included in the quoted theory
error. This constitutes the most precise measurement of $R_{D\pi}$ to date.
The measured value of ${\cal B}\left(\Bzb\to\Ds\Km\right)$ can be understood
in terms of a pure $W$-exchange contribution, which is in agreement with our
recent measurement of $\Bzb\to\Dss\Km$~\cite{nikhil}.

To conclude, using a data sample of $\bbpairs$ $\BB$ pairs collected by
Belle, we report the most precise measurement of branching fractions for
the $\Bz\to\Ds\pim$ and $\Bzb\to\Ds\Km$ decays. This improves the precision
of the parameter $R_{D\pi}$, and thus will also improve determinations of
the UT angle $\phi_3$ from $\CP$ violation measurements in $\Bz\to\Dmp\pipm$
decays. One can use the $\Bz\to\Ds\pim$ result to calculate the CKM matrix
element $|V_{ub}|$ following the prescription laid out in Ref.~\cite{vub}.
Our results supersede the previous Belle measurement~\cite{belleold}.

We thank the KEKB group for the excellent operation of the
accelerator, the KEK cryogenics group for the efficient
operation of the solenoid, and the KEK computer group and
the National Institute of Informatics for valuable computing
and SINET3 network support.  We acknowledge support from
the Ministry of Education, Culture, Sports, Science, and
Technology (MEXT) of Japan, the Japan Society for the 
Promotion of Science (JSPS), and the Tau-Lepton Physics 
Research Center of Nagoya University; 
the Australian Research Council and the Australian 
Department of Industry, Innovation, Science and Research;
the National Natural Science Foundation of China under
contract No.~10575109, 10775142, 10875115 and 10825524; 
the Ministry of Education, Youth and Sports of the Czech 
Republic under contract No.~LA10033 and MSM0021620859;
the Department of Science and Technology of India; 
the BK21 and WCU program of the Ministry Education Science and
Technology, National Research Foundation of Korea,
and NSDC of the Korea Institute of Science and Technology Information;
the Polish Ministry of Science and Higher Education;
the Ministry of Education and Science of the Russian
Federation and the Russian Federal Agency for Atomic Energy;
the Slovenian Research Agency;  the Swiss
National Science Foundation; the National Science Council
and the Ministry of Education of Taiwan; and the U.S.
Department of Energy.
This work is supported by a Grant-in-Aid from MEXT for 
Science Research in a Priority Area (``New Development of 
Flavor Physics''), and from JSPS for Creative Scientific 
Research (``Evolution of Tau-lepton Physics'').


\begin{thebibliography}{99}

\bibitem{ckm}
  N. Cabibbo, Phys.\ Rev.\ Lett. {\bf 10}, 531 (1963);
  M. Kobayashi and T. Maskawa, Prog.\ Theor.\ Phys. {\bf 49}, 652 (1973). 

\bibitem{ut}
  {\it CP Violation}, edited by C. Jarlskog, Advanced Series on Directions in High
  Energy Physics (World Scientific, Singapore, 1989), Vol. 3.

\bibitem{rdpi}
  I. Dunietz and R.~G. Sachs, Phys.\ Rev.\ D {\bf 37}, 3186 (1988);
  I. Dunietz, Phys.\ Lett.\ B {\bf 427}, 179 (1998);
  D.~A. Suprun, C.~W. Chiang, and J.~L. Rosner, Phys.\ Rev.\ D {\bf 65}, 054025 (2002).

\bibitem{dpicpv}
  F.~J. Ronga {\it et al.} (Belle Collaboration), Phys.\ Rev.\ D {\bf 73}, 092003 (2006);
  B. Aubert {\it et al.} (BaBar Collaboration), Phys.\ Rev.\ D {\bf 73}, 111101 (2006).

\bibitem{bellephi1}
  K.~F. Chen (Belle Collaboration), Phys.\ Rev.\ Lett. {\bf 98}, 031802 (2007).

\bibitem{babarbeta}
  B. Aubert (BaBar Collaboration), Phys.\ Rev.\ D {\bf 79}, 072009 (2009).

\bibitem{vub}
  D. Choudhury, D. Indumati, A. Soni, and S. Uma Sankar, Phys.\ Rev.\ D {\bf 45}, 217 (1992);
  C.~S. Kim, Y. Kwon, J. Lee, and W. Namgung, Phys.\ Rev.\ D {\bf 63}, 094506 (2001);
  C.~S. Kim and Y. Li, arXiv:1007.2291v2.

\bibitem{dsk1}
  B. Blok, M. Gronau, and J.~L. Rosner, Phys.\ Rev.\ Lett. {\bf 78}, 3999 (1997);
  D. Du, L. Guo, and D. Zhang, Phys.\ Lett.\ B {\bf 406}, 110 (1997);
  C.~K. Chua, W.~S. Hou, and K.~C. Yang, Phys.\ Rev.\ D {\bf 65}, 096007 (2002);
  C.~W. Chiang, Z. Luo, and J.~L. Rosner, Phys.\ Rev.\ D {\bf 66}, 057503 (2002);
  N. Mahajan, Phys.\ Lett.\ B {\bf 634}, 240 (2006).

\bibitem{dsk2}
  C.~D. Lu and K. Ukai, Eur.\ Phys.\ J.\ C {\bf 28}, 305 (2003);
  M. Gronau and J.~L. Rosner, Phys.\ Lett.\ B {\bf 666}, 185 (2008).

\bibitem{belleold}
  P. Krokovny {\it et al.} (Belle Collaboration), Phys.\ Rev.\ Lett. {\bf 89}, 231804 (2002).

\bibitem{babarold}
  B. Aubert {\it et al.} (BaBar Collaboration), Phys.\ Rev.\ D {\bf 78}, 032005 (2008).

\bibitem{kekb}
  S. Kurokawa and E. Kikutani, Nucl.\ Instrum.\ Methods\ Phys.\ Res., Sect.\ A {\bf 499}, 1 (2003), and other papers included in this volume.

\bibitem{belledetector}
  A. Abashian {\it et al.} (Belle Collaboration),
  Nucl.\ Instrum.\ Methods\ Phys.\ Res., Sect.\ A {\bf 479}, 117 (2002).

\bibitem{charge}
  Unless explicitly stated otherwise, inclusion of charge conjugate processes is
  implied throughout the paper.

\bibitem{PDG}
  K. Nakamura {\it et al.} (Particle Data Group), J.\ Phys.\ G {\bf 37}, 075021 (2010).

\bibitem{selks}
  K.~F. Chen {\it et al.} (Belle Collaboration), Phys.\ Rev.\ D {\bf 72}, 012004 (2005).

\bibitem{fox}
  G.~C. Fox and S. Wolfram, Phys.\ Rev.\ Lett. {\bf 41}, 1581 (1978).
  The modified moments used in this paper are described in
  S.~H. Lee {\it et al.} (Belle Collaboration), Phys.\ Rev.\ Lett. {\bf 91}, 261801 (2003).

\bibitem{MCsim}
  For MC event generation, the \evtgen\ package, described in D.~J. Lange, Nucl.\ Instrum.\
  Methods\ Phys.\ Res., Sect.\ A {\bf 462}, 152 (2001), is used, while the detector
  response is simulated using \geant, described in R. Brun {\it et al.}, CERN Report No.
  CERN-DD-EE-84-1, 1987.

\bibitem{dsratio}
  E. Follana {\it et al.} (HPQCD and UKQCD Collaborations),
  Phys.\ Rev.\ Lett. {\bf 100}, 062002 (2008).

\bibitem{maxbaak}
  M.~A. Baak, Ph.D. thesis, Vrije University [SLAC Report No. SLAC-R-858, 2007].

\bibitem{nikhil}
  N.~J. Joshi {\it et al.} (Belle Collaboration), Phys.\ Rev.\ D {\bf 81}, 031101 (2010).

\end{thebibliography}
\end{document}